\documentclass[10pt, conference, compsocconf]{IEEEtran}
\usepackage{booktabs} 
\usepackage{amssymb}
\usepackage{xcolor}
\usepackage{comment}

\ifCLASSINFOpdf
   \usepackage[pdftex]{graphicx}
   \graphicspath{{../pdf/}{../jpeg/}}
   \DeclareGraphicsExtensions{.pdf,.jpeg,.png}
\else
\fi
\usepackage[cmex10]{amsmath}
\usepackage{algorithmic}
\usepackage[font=footnotesize]{subfig}
\begin{document}
%
\title{Adaptive Pattern Matching with Reinforcement Learning for Dynamic Graphs}




%
\author{\IEEEauthorblockN{Hiroki Kanezashi\IEEEauthorrefmark{1}\IEEEauthorrefmark{2},
Toyotaro Suzumura\IEEEauthorrefmark{2}\IEEEauthorrefmark{3},
Dario Garcia-Gasulla\IEEEauthorrefmark{3},
Min-hwan Oh\IEEEauthorrefmark{4}\IEEEauthorrefmark{2}
and Satoshi Matsuoka\IEEEauthorrefmark{5}\IEEEauthorrefmark{1}}
\IEEEauthorblockA{\IEEEauthorrefmark{1}Tokyo Institute of Technology\\
2-12-1, Ookayama, Meguro-ku, Tokyo, Japan 152-8550\\ Email: kanezashi.h.aa@m.titech.ac.jp}
\IEEEauthorblockA{\IEEEauthorrefmark{2}IBM T.J. Watson Research Center\\
1101 Kitchawan Rd, Yorktown Heights, New York, 10598}
\IEEEauthorblockA{\IEEEauthorrefmark{3}Barcelona Supercomputing Center\\
29-31, Carrer de Jordi Girona, Barcelona, Spain, 08034}
\IEEEauthorblockA{\IEEEauthorrefmark{4}Columbia University\\
500 W 120th St.  New York, NY 10027}
\IEEEauthorblockA{\IEEEauthorrefmark{5}RIKEN Center for Computational Science\\
7-1-26, Minatojima-minamimachi, Chuo-ku, Kobe, Hyogo, Japan, 650-0047}}


\maketitle

\begin{abstract}
Graph pattern matching algorithms to handle million-scale dynamic graphs are widely used in many applications such as social network analytics and suspicious transaction detections from financial networks.
On the other hand, the computation complexity of many graph pattern matching algorithms is expensive, and it is not affordable to extract patterns from million-scale graphs. Moreover, most real-world networks are time-evolving, updating their structures continuously, which makes it harder to update and output newly matched patterns in real time.
Many incremental graph pattern matching algorithms which reduce the number of updates have been proposed to handle such dynamic graphs. However, it is still challenging to recompute vertices in the incremental graph pattern matching algorithms in a single process, and that prevents the real-time analysis.
We propose an incremental graph pattern matching algorithm to deal with time-evolving graph data and also propose an adaptive optimization system based on reinforcement learning to recompute vertices in the incremental process more efficiently. Then we discuss the qualitative efficiency of our system with several types of data graphs and pattern graphs.
We evaluate the performance using million-scale attributed and time-evolving social graphs. Our incremental algorithm is up to 10.1 times faster than an existing graph pattern matching and 1.95 times faster with the adaptive systems in a computation node than naive incremental processing.
\end{abstract}


%
\IEEEpeerreviewmaketitle

\section{Introduction}

\subsection{Background}
Graph pattern matching is a fundamental graph data processing method. It extracts subgraphs with some restrictions such as topologies or vertex and edge attributes represented as a small graph \textbf{(query pattern graph)} from a large-scale graph \textbf{(input data graph)}. Graph pattern matching algorithms are widely used in many applications, such as social network analysis \cite{fan2012graph} and suspicious financial transaction detection \cite{michalak2011graph}.
In social network analysis, for example, graph pattern matching algorithms enable the extraction of particular organizations and clubs of members and relationships with typical roles. In suspicious financial transaction detection, it allows frequent screenings of bank accounts involving suspicious transactions.
These networks typically used in these real-world applications are usually million-scale, and their structures update frequently.

\subsection{Problem Definition}
There exist significant challenges while performing graph analytics with time-evolving networks.

First of all, real-world graphs update their topologies and attributes of their vertices and edges frequently. Many graph analytics needs to follow the update, but it is hard to re-compute for the entire graph after every change to avoid missing newly matched patterns.

It is also difficult to apply a pattern-matching algorithm to full graph structures with million edges because it has at least polynomial time complexity. In particular, the computation complexity of subgraph isomorphism is NP-complete \cite{lewis1983computers}. In order to run graph pattern matching in realistic execution time, we need to adjust the computation area and adopt an approximate algorithm with less computation complexity.

\subsection{Requirements of Graph Pattern Matching}
\label{sec:req}

Optimization methods for graph pattern matching algorithms have been proposed to deal with these problems. However, some of them are not suitable for real-world network analysis because they do not support topological matching and vertex or edge attribute matching. There are pattern matching algorithms which support exact topological matching, but it takes a long time to update results incrementally even it outputs few matched patterns with the time constraint.

To propose a graph pattern matching method addressing these problems, we define the following requirements of our proposed graph pattern matching method. These requirements are fundamental to conduct flexible and real-time graph pattern matching for real-world applications.

\textbf{Subgraph isomorphism:}
There are graph pattern matching algorithms such as graph simulations which do not consider the topological matching in order to reduce the computational cost of subgraph isomorphism. However, for many applications, such topological restrictions restrict the patterns that are worth exploring.

\textbf{Attributes of vertices and edges:}
Vertices and edges in real-world graph data usually have attributes such as labels and properties. In financial transaction data, for example, vertices (bank accounts) have account ID, customer name, or account type while edges (remittances) have transaction date and amount. Users extract patterns with restrictions of such attributes.

To address this issue, we add to our graph pattern matching method a requirement to support the label and property filtering for both vertices and edges. In other words, when the vertices and edges in a query graph have labels and properties, we require that the corresponding vertices and edges in the matched subgraphs also have labels and properties.

\textbf{Approximate patterns should also be found:}
As we mentioned at the first requirement, topological restrictions are necessary for many graph pattern matching applications.
On the other hand, some applications favor outputs of topologically approximate patterns. For example, in a suspicious transaction pattern detection from financial transaction graphs, it is unrealistic to prepare all criminal transaction patterns and match them exactly. There might be potentially suspicious transaction patterns similar to obviously suspicious patterns. In this situation, our proposal allows to topologically approximate patterns as well as exact patterns.

\subsection{Research Challenges and Contributions}
Many matching problem for large-scale and time-evolving graphs have been proposed. Previous work proposed in \cite{fan2012graph, abdelhamid2017incremental, wickramaarachchi2016distributed, han2013turbo} updates the results partially by the update of input data graph structures and attributes.
However, the efficiency of the incremental process depends on the types of the input data graphs and query pattern graphs. 

We define the following research challenges.
\begin{itemize}
    \item How can we reduce computation costs of an approximate subgraph isomorphism algorithm for dynamic graphs? Can we update the partial result of graph pattern matching efficiently?
    \item Can we adjust the incremental processing from the type of input data graphs and query pattern graphs automatically?
    \item Can we make the optimization method applicable to various input graphs and query patterns?
\end{itemize}

In this work, we propose an adaptive and incremental graph pattern matching algorithm with reinforcement learning.

First, we propose an incremental graph pattern matching \textbf{(IGPM)} algorithm based on an existing approximate graph pattern matching algorithm named G-Ray \cite{tong2007fast}. G-Ray supports a ``best-effort'' topological pattern matching with reduction of the computation complexity. It has core functions to search for the best matched vertices and edges. We extend the incremental version of these functions according to the type of graph updates: edge additions, edge removals and vertex label updates.

We also propose ``partial execution manager" \textbf{(PEM)} to choose subgraphs for re-computation by our incremental graph pattern matching algorithm. The best matching vertices set depends on several factors including the number of vertices and edges in the input data graph and required time limit. It partitions input data graphs into small clusters and output vertices in the updated clusters.
The PEM component has a reinforcement learning sub-component to receive feedback such as elapsed time and the number of matched patterns. This information is used to adjust these parameters of the graph clustering.

In summary, our technical contributions are as follows.
\begin{enumerate}
    \item We propose an incremental graph pattern matching \textbf{(IGPM)} algorithm based on an approximate and fast graph pattern matching algorithm to reduce re-computation cost.
    \item We propose a performance optimization component named partial execution manager \textbf{(PEM)} for \textbf{IGPM} to output the best vertex set for re-computations in the incremental processing.
    \item We show the efficiency of our algorithm and system using some input data graph and query pattern graph types commonly used in real-world graph data sets and pattern matching applications.
    \item We evaluate the performance of \textbf{IGPM} and \textbf{PEM} using real-world time-evolving graph data sets and several types of query pattern graphs in a multi-core computation node.
\end{enumerate}

\section{Related Work}

Previous works proposed graph pattern matching algorithms which deal with large-scale graphs. Some of them relax topological restrictions of subgraph isomorphism due to the high computation complexity, and others apply clustering algorithms to input data graphs to prune unnecessary re-computation areas.

\subsection{Graph Pattern Matching for Large-scale Graphs}
Graph pattern matching algorithms are categorized to \textit{subgraph isomorphism} \cite{cordella2004sub} and \textit{graph simulation} \cite{henzinger1995computing}. The main difference is whether it requires topological matching including edges or it considers only corresponding vertices.
Subgraph isomorphism extracts subgraphs with the same graph topology as the query graph. On the other hand, graph simulation relaxes the topological restriction and only take vertex matching into considerations.

Because most real-world applications require the topological constraints, subgraph isomorphism is a better fit as it can extract specific patterns by specifying topological constraints.
However, because the computational complexity of subgraph isomorphism problem is NP-complete \cite{lewis1983computers}, many relaxed pattern matching algorithms have been proposed \cite{fan2012graph} \cite{fan2011incremental} \cite{li2017relaxing}.

The computation cost is linear with the number of matched patterns; hence it is unrealistic to extract all possible patterns. \cite{stotz2009incremental} \cite{sun2012efficient} \cite{fan2014querying} \cite{nasir2017fully} extract only top-k patterns. These methods do not guarantee to extract all possible patterns, but they consider the topological matching to a certain degree.

Some graph pattern matching algorithms for large-scale and attributed graph are proposed with the graph data optimization techniques such as query graph decomposition \cite{sun2012efficient} \cite{choudhury2013fast} \cite{yang2016fast} and input data graph compression \cite{fan2012query} to reduce the computation costs keeping the output qualities.

\subsection{Incremental Graph Pattern Matching}

Table \ref{tb:relatedwork} represents the state-of-art incremental graph pattern matching algorithms recently proposed for large-scale graph data. It also represents the compatibility with the requirements we defined in Section \ref{sec:req}. Columns "ISO" (subgraph isomorphism), "Attr" (vertex and edge attributes), "Time" (real-time processing) and "Approx" (approximate matching) indicate whether these methods fulfill each requirement.
Our system with the incremental graph pattern matching algorithm and an optimization component for incremental process supports all four features.

\begin{table*}[htbp]
\centering
\begin{tabular}{|l|l|l|l|l|l|} 
 \hline
 Paper & Method & ISO & Attr & Time & Approx \\ \hline
 IncGM+ \cite{abdelhamid2017incremental} & Compute only fringe subgraphs & \checkmark & \checkmark &  & \checkmark \\ \hline
 D-ISI \cite{wickramaarachchi2016distributed} & Distributed graph pruning & \checkmark & \checkmark &  & \checkmark \\ \hline
 Bounded Simulation \cite{fan2012graph} & Query preserving graph compression &  &  & \checkmark & \checkmark \\ \hline
 Turbo ISO \cite{han2013turbo} & Region exploration and COMB/PERM & \checkmark & \checkmark & \checkmark & \\ \hline
 Our Method & Adaptive and approximate IGPM and PEM & \checkmark & \checkmark & \checkmark & \checkmark \\ \hline
\end{tabular}
\caption{The state-of-art incremental graph pattern matching methods}
\label{tb:relatedwork}
\end{table*}

In \cite{abdelhamid2017incremental}, Abdelhamid et al. proposed IncGM+, a heuristic subgraph isomorphism algorithm for large evolving graphs. It separates the input data graph into frequent and infrequent updated subgraphs and prunes the update area by adjusting the boundary subgraphs named "fringe". They also proposed the optimization of subgraph embeddings in the fringe to reduce the memory overhead for storing fringe subgraphs.
It reduced the computation time keeping the small memory overhead, but some datasets did not improve the performances because of too much graph pruning were performed.

In \cite{wickramaarachchi2016distributed}, Wickramaarachchi et al. proposed a distributed graph pruning algorithm for dynamic graphs (\textit{D-IDS}) and a distributed incremental subgraph isomorphism algorithm (\textit{D-ISI}). \textit{D-IDS} limits the search space for subgraph isomorphism with small diameter graphs. This method can extract exact patterns with lower latency in small-diameter input graphs, but the latency is much longer in the large-diameter graphs.

Fan \cite{fan2012graph} proposed a notion of graph pattern matching the revised incremental version of "Bounded Simulation" \cite{fan2010graph} for social network data with optimizations by query preserving graph compression and distributed graph pattern matching. 
However, even though it considers the connectivity of the two matched vertices connected each other, the limitation is still permissive, and it does not evaluate the similarity between the query pattern and extracted patterns.

\section{Proposed Method}
To process incremental graph pattern matching efficiently, we propose an adaptive graph pattern matching algorithm named IGPM-PEM, with incremental graph pattern matching (IGPM) algorithm and partial execution manager (PEM) to re-compute the updated subgraphs.

In this section, we present our proposed IGPM algorithm and PEM for re-computation optimization. First, we introduce an incremental and approximate subgraph isomorphism algorithm based on G-Ray \cite{tong2007fast} algorithm. Next, we introduce PEM to optimize the IGPM consists of reinforcement learning and graph clustering components. Finally, we discuss the efficiency of several types of input data graph and query pattern graph qualitatively.

\begin{figure}[htbp]
    \centering
    \includegraphics[width=0.45\textwidth]{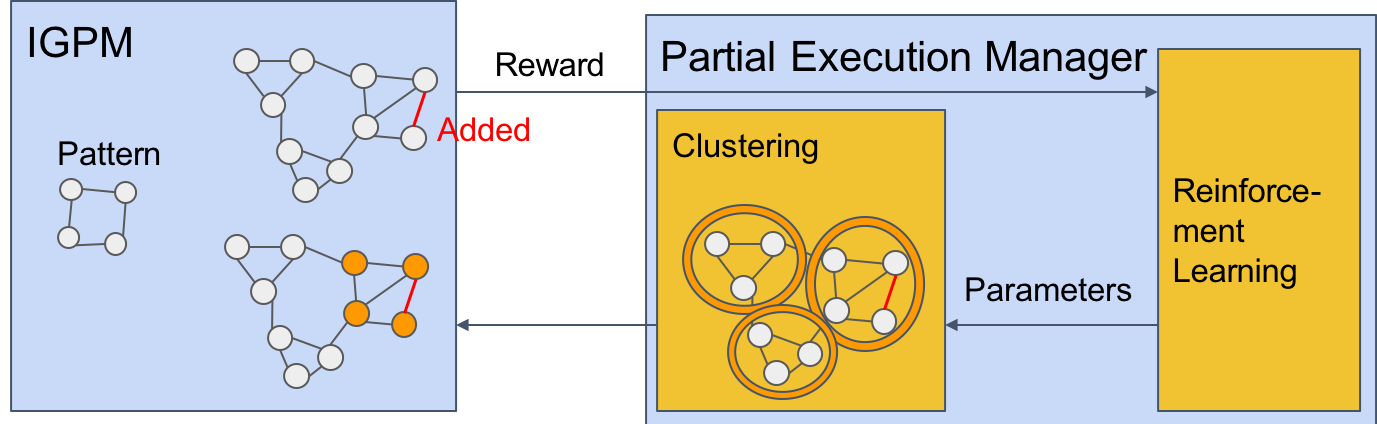}
    \caption{Components of our adaptive IGPM system}
    \label{fig:system}
\end{figure}

Figure \ref{fig:system} represents our system with IGPM-PEM component. It iterates our incremental graph pattern matching algorithm, reinforcement learning, and graph clustering processes to optimize the performance of graph pattern matching.

First, when some topological or attribute updates occur in the input data graph such as edge additions or vertex label changes in the IGPM component, the PEM component inputs the accumulated features of the data graph such as the number of vertices and edges. It also inputs the elapsed time of the IGPM phase and the number of extracted patterns and their similarities as rewards of the reinforcement learning.

The reinforcement learning component computes the best granularity of subgraphs for the re-computation process from the reward and input parameters such as input graph data size. It sends the minimum number of the subgraph size to the graph clustering component.

The graph clustering component partitions the input data graph into the clusters with the appropriate size. It repeats the Louvain method \cite{blondel2008fast} until these partitioned clusters cannot be divided further or the size of clusters is smaller than the parameter defined at the reinforcement learning component.
If a cluster has at least one updated graph element such as an added edge, all vertices in the cluster will be subject to the re-computation of IGPM in the next iteration. Finally, PEM sends the resultant vertex set to the IGPM component.

\subsection{Approximate Graph Pattern Matching}
Previously we mentioned the requirements of graph pattern matching that topologically matches exact patterns and similar patterns should be extracted from the input data graph.
For this reason, we extended an existing approximate subgraph isomorphism algorithm which extracts both exact and approximate patterns.

G-Ray \cite{tong2007fast} is an approximate subgraph isomorphism algorithm which extracts subgraphs matched with the desirable patterns which are ``best-effort results''. That means each vertex in the subgraph corresponds to a vertex in the query graph, and the connectivity is maintained even when the exact pattern does not exist in the input data graph.
Additionally, the computation cost of the G-Ray is linear concerning the size of the data graph compared to the polynomial complexity of the exact subgraph isomorphism algorithms.

We adopt G-Ray as a baseline of our method since it has following properties, which satisfy part of the requirements we defined.

\textbf{Approximate subgraph isomorphism:} It exacts the most proper subgraphs for the query pattern as subgraph isomorphism algorithms. When it finds matched edges directly connecting two matched vertices, it will choose the most suitable one with the highest goodness score.

\textbf{Considers vertex label matching:} It finds each matched vertex from input data graph with the same vertex label as in the query pattern graph.

\textbf{Extract a specified number of best-effort subgraphs:} It extracts similar patterns as well as exact patterns by the best-effort matching process. Users can specify the number of patterns to extract.

\textbf{Linear computation complexity:} Graph pattern matching algorithms based on subgraph isomorphism are NP-complete, which is prohibitive for large-scale graphs. On the other hand, the computation complexity of G-Ray algorithm is the linear of the number of vertices in the data graph.

G-Ray has the following three core functions to find the best matched first vertex named ``seed'' vertex, neighbor vertices of the previously matched vertices, and edges connecting these matched vertices.

\textbf{Seed-finder:} It finds the most suitable vertex (seed vertex) from the input data graph, producing the largest goodness function score corresponding to a query graph with the same label.

\textbf{Neighbor-expander:} It finds the best matching neighbor vertex from the input data graph of the previously matched vertex, producing the best goodness function from the input graph to indicate the closeness of two vertices.

\textbf{Bridge:} It finds the best path (one or more connecting edges) between these two matched vertices with the derivative of the EXTRACT \cite{tong2006center} algorithm, and connects them from the input data graph with intermediate edges.

The two matched vertices are not always neighboring each other because there are no such directly connecting edges between these vertices which exist in the query graph. Even in this case, G-Ray finds the best (usually the shortest) paths in the bridge function and output the "best-effort" approximate query.

\subsection{Incremental Graph Pattern Matching}
To execute IGPM efficiently of time-evolving graphs, we design the incremental version of the G-Ray algorithm by extending these three core functions of the original G-Ray. Figure \ref{alg:igpm-all} shows the algorithm of IGPM.


\begin{figure}[htb]
\begin{algorithmic}[1]
\REQUIRE The input data graph $G$, the query pattern graph $H_q$, vertices list $V_l$ from updated edges
\ENSURE Updated subgraphs $H_l$
\FOR{$v \in V_l$}
    \STATE Update all RWR values $r_{v,u}$ from $v$ to reachable vertices $u$
\ENDFOR
\FOR{$v \in V_l$}
    \STATE $seed \leftarrow v$
    \REPEAT
        \FOR{$u \in neighbors(v)$}
            \STATE Compute goodness function $g_u$ from $r_{v,u}$
        \ENDFOR
        \STATE $u' = argmax_{u} (g_u)$
        \STATE Update the best path by \textbf{bridge} algorithm with $r_{v,u}$
        \STATE Mark the edge $(v, u)$ in $H_q$ as "processed"
    \UNTIL{All nodes in $H_q$ are "processed"}
    \STATE Add the extracted pattern to $H_l$
\ENDFOR
\end{algorithmic}
\caption{IGPM Algorithm}
\label{alg:igpm-all}
\end{figure}

The incremental extension of seed-finder function performs different procedures by the change of graph structures or attributes. It re-computes the RWR scores and goodness scores of vertex pairs should be directly connected. Then, it iterates the incremental extension of ``neighbor-expander'' and ``bridge'' functions for each vertex as a ``seed'' connected to the added or removed edges.

The incremental extension of neighbor-expander function updates the best-matched neighbor vertex of the previously matched vertex. When the added edge is connected to the "seed" vertex, it re-computes the goodness function of the seed vertex and all the neighboring vertices. If some different neighbor vertices produce the higher goodness function score than that score of the current neighbor vertex, replace it with the new vertex with the highest score.

When the seed vertex or the neighbor vertices are updated after applying the incremental seed-finder or incremental neighbor-expander, it needs to re-compute the original bridge function with these updated matching vertices. If these origin and destination vertices the same as in the previous state, it only checks the updates of intermediate vertices and edges, and re-computes and reconnects the path partially. When an edge is added to the input data graph, it extracts all existing paths passing the source and destination vertices and re-computes goodness score and these intermediate paths between them if needed.

\subsection{Partial Execution Manager}
To execute our IGPM algorithm with real-world dynamic graphs more effectively, we also propose \textbf{Partial Execution Manager (PEM)}, an adaptive vertex assignment component for dynamic graphs. PEM computes a vertex set for re-computation from the current graph states and parameters.

It consists of two sub-components: the graph clustering component and the reinforcement learning component. The reinforcement learning component adjusts the parameter of the graph clustering component which determines the area of subgraphs to extract vertices for re-computations through the Louvain method.

Figure \ref{alg:igpm-pem} shows the optimized IGPM algorithm with PEM. Lines from 7 to 20 represents the PEM process: reinforcement learning and graph clustering are performed at line 7 and 13.

\begin{figure}[htb]
\begin{algorithmic}[1]
\REQUIRE The input data graph $G$, the query pattern graph $H_q$, initial community size $c$, total number of steps $T$
\ENSURE Subgraph patterns $H_l$
\STATE $step \leftarrow 0$
\REPEAT
    \STATE $V_l \leftarrow$ vertices from updated edges at $step$
    \STATE $H_l \leftarrow IGPM(G, H_q, V_l)$
    \STATE $t \leftarrow$ elapsed time of $IGPM$
    \STATE $x \leftarrow$ number of vertices and edges of $G$
    \STATE $y \leftarrow$ reinforcement learning with input $x$ and reward $\frac{1}{t}$
    \IF{$y == 0$}
        \STATE $c \leftarrow c - 1$
    \ELSE
        \STATE $c \leftarrow c + 1$
    \ENDIF
    \STATE $communities \leftarrow$ clustering with recursive Louvain from $G$ into communities with size $c$
    \STATE $l \leftarrow$ vertices from updated edges at $step + 1$
    \STATE $V_l = list()$
    \FOR{$com \in communities$}
        \IF{$com$ and $l$ has common vertices}
            \STATE Add all vertices in $com$ to $V_l$
        \ENDIF
    \ENDFOR
\UNTIL{$step > T$}
\end{algorithmic}
\caption{Optimized IGPM with Partial Execution Manager}
\label{alg:igpm-pem}
\end{figure}

\subsubsection{Re-computation Vertices Extraction}
In incremental graph pattern matching, we need to re-compute neighboring vertices of the updated area to find new patterns because even an edge update will affect the result of RWR in the neighboring vertices. On the other hand, it is inefficient to re-compute all unaffected vertices to find patterns with only a few vertex or edge updates.
To extract re-computation vertices, determining the subgraph areas is important. PEM adjusts the area adaptively using graph clustering and reinforcement learning to extract new patterns.

When too many vertices updated to be re-computed all vertices, PEM extracts the minimum number of vertices from fine-grained subgraphs. The reinforcement learning component outputs the smaller number of community size for the graph clustering component to output the fine-grained vertex set.

Then, the graph clustering component repeats the Louvain methods recursively until each size of clusters is less than the specified size threshold and outputs small communities. Finally, PEM extracts vertices from these communities, and output all vertices in each community if at least one of vertices in the community also belongs to the updated vertex set.

\subsubsection{Data Graph Clustering}
PEM has a graph clustering component to divide input data graph into small clusters and extract vertices for re-computation from the several clusters where edge updates occur.

We use the Louvain method \cite{blondel2008fast} as a clustering algorithm. It is a heuristic community extraction method with optimizing modularity. It iterates changes of communities locally to maximize modularity and aggregates these communities.

The main reason we use the Louvain method as clustering in PEM is that we can adjust the size of communities by the number of iterations.
When it has insufficient time and does not want so many patterns, it partitions the graph into fine-grained clusters and output minimum vertices for re-computations.
Moreover, it can handle larger (million-scale) graphs in the reasonable computation cost and the additional time for the graph clustering process itself is relatively small.

\subsubsection{Reinforcement Learning}
There are many factors to find the best subgraph area for re-computations, such as the input data graph and query pattern graph, updated edges and so on.
However, it is difficult to find out the best parameter because there is no answer data (the best parameter) for unknown data sets (input data graph and query pattern graph), and it is also hard to learn parameters for all combinations of data sets.

To avoid the problem, PEM has reinforcement learning component to adjust the input parameters of graph clustering component automatically. It updates its learning model to output the best parameter during the iteration of IGPM and PEM process, so we do not have to try many parameters and formulas of the model manually. 

We use a Deep-Q Network (DQN) \cite{mnih2015human} for our reinforce learning model. The observation of the reinforcement learning agent, i.e., input layer of this DQN, are two-dimensional real-valued variables: the density of input data graph (a proportion of the number of edges out of the number of vertices) and the proportion of the affected communities. The DQN has two hidden layers with four units, and each layer is fully connected. The output layer to determine an action has two units. The action is to increment or to decrement the threshold of minimum community size.

\subsection{Qualitative Discussion of Graph Types}
\label{sec:discussion}
In graph pattern matching applications with real-world data set, users may want to try several kinds of queries according to the kind of applications or data sets.
In this section, we discuss the combinations of input data graph and query pattern graph types in detail to prove the validity of our method using PEM qualitatively.

\subsubsection{Input Data Graphs}

Considering our method is used in primary large-scale real-world applications, first we discuss the validity for the following five types of input data graph and kind of clustering algorithms.

\begin{enumerate}
    \item Scale-free graph
    \item Random graph
    \item Sparse graph (with isolated edges)
    \item Sparse graph (with dense subgraphs)
    \item Dense graph
\end{enumerate}

\textbf{Scale-free graph:}
Scale-free graphs are considered as the representation of real-world networks. They have a few hub vertices and many leaf vertices and usually appear in many real-world domains such as social networks.
However, clustering algorithms which maximize the modularity are not suitable for such graphs. Since the degree distribution is unbalanced, clustering algorithms usually output unbalanced clusters, and it is difficult to adjust these sizes. The cluster size depends on the degree of hub vertices.

\textbf{Random graph:}
Since the degree distributions of random graphs are relatively decentralized, compared to scale-free graphs, the result of clustering can be more balanced and flexible.

\textbf{Sparse graph with isolated edges:}
Sparse graphs appear as the initial states of real-world networks such as social network and bank transactions.
These type of graphs usually have many isolated vertices, edges and tiny components, which should be excluded in advance to apply clustering operations to large components.

\textbf{Sparse graph with dense subgraphs:}
When a sparse graph is growing to a certain degree, it usually has dense subgraphs. Such a graph is the most suitable graph type for clustering. It is also suitable for dense, cycle query graphs, but not for line query graphs. Line patterns found in previous steps will be disconnected by clustering.

\textbf{Dense graph:}
Since there are many edges compared to the number of vertices, it has high computation costs for re-computation of G-Ray, particularly about neighbor-expander and bridge functions.
Also, because many edges are disconnected by a clustering algorithm such as Louvain method, IGPM may miss many patterns. A minimum-cut clustering algorithm such as METIS \cite{karypis1998fast} may provide better results for such graph type. Such clustering algorithms will output unbalanced clusters and results to less efficiency, but it minimizes the number of new missed patterns.

\subsubsection{Query Pattern Graphs}
\label{sec:query}

Not only input data graphs but query pattern graphs have many structure types, and the compatibility depends on the structure of input data graphs.
To explain our adaptive IGPM system is effective for various query pattern graphs as well as input data graphs, we need to confirm the efficiency of significant types of query patterns one by one.

\begin{enumerate}
    \item Star query
    \item Cycle query
    \item Dense query
\end{enumerate}

Star query graphs consist of a single high-degree "hub" vertex and many "leaf" vertices which directly connect to the hub vertex. Such hub vertices exist in the scale-free, random and dense graphs or clusters, so it can be applied these input data graphs.
In our IGPM algorithm, the "seed" vertex of the G-Ray is always "hub" vertex because high-degree vertices derived from RWR usually have high goodness score in the G-Ray algorithm.
When we use star query graphs in our IGPM derived from the G-Ray algorithm, we have only to find and re-compute subgraphs with hub vertices with the similar degree to the hub vertices in query graphs.
That means it has only to re-compute hub vertices in many cases because other leaf vertices can never be "seed" vertices.

Cycle query graphs are also sequences of vertices and edges such as line query graphs, but the structure is closed.
Such patterns are used for some specific applications. For example, they can detect transaction loops involving money laundering in bank transaction graphs and suspicious corporation relationships in finance risk detection applications.
The diameter of the cycle query is relatively large so that when graph clustering process divides dense input data graphs and many edges are disconnected, some pattern subgraphs crossing clusters will be missed.
In dense graphs, IGPM of cycle query graph is difficult because it needs to update long matched cycles.
However, there are some hints to choose vertices for re-computation. First, all vertices must be at least 2-degree. Second, hub vertices are likely to be part of cycles. By the hints, we can set the priorities to high-degree vertices for re-computation.

Dense query graph means a query graph with many edges compared to the number of vertices because our IGPM algorithm needs to repeat neighbor-expander and bridge functions many times. One or more vertices in a dense graph have high degree and usually connected to all other vertices.
There are some features of the pattern matching using dense query graphs. First, it can match only with dense graphs (or dense subgraphs). Moreover, such as cycle query graphs, when many edges are disconnected by a clustering, some patterns will be missed.

\subsubsection{Summary of the qualitative discussions}
Based on these above discussions, the compatibility of input data graphs, graph clustering algorithms and query pattern graphs is the Table \ref{tb:compatibility}. "Min-Cut" or "Max-Mod" in the "Clustering" column means that minimum cut (e.g. METIS \cite{karypis1998fast}) or modularity maximization (e.g. Louvain \cite{blondel2008fast}) is preferable.

\begin{table}[htb]
\centering
\begin{tabular}{|l|l|l|l|l|} 
 \hline
 Input Data Graph & Clustering & Star & Cycle & Dense \\ \hline
 Scale-free & Min-Cut & \checkmark & \checkmark & \\ \hline
 Random & Max-Mod & \checkmark & \checkmark & \\ \hline
 Sparse, isolated edges & Max-Mod &  & \checkmark & \\ \hline
 Sparse, dense clusters & Max-Mod & \checkmark & \checkmark & \checkmark \\ \hline
 Dense & Min-Cut & \checkmark &  & \checkmark \\ \hline
\end{tabular}
\caption{Compatibility of input data and query pattern graph types and clustering methods}
\label{tb:compatibility}
\end{table}

\section{Evaluation}
To evaluate the effect of our IGPM-PEM system, we conducted performance evaluations with real-world graph datasets. We also validate the qualitative discussion in section \ref{sec:discussion}.

\subsection{Implementation}
\begin{figure}[htbp]
    \centering
    \includegraphics[width=0.5\textwidth]{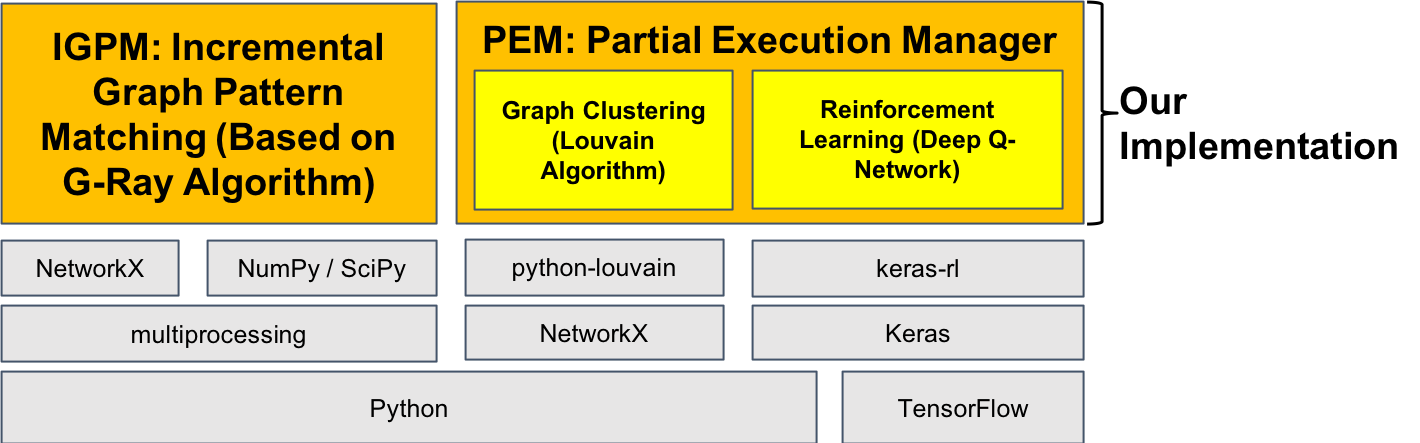}
    \caption{Software stack of IGPM and PEM}
    \label{fig:stack}
\end{figure}

We implemented batch and incremental graph pattern matching algorithm, Partial Execution Manager in Python 2.7.13 and the NetworkX 2.1 package for graph processing. As the reinforcement learning framework in PEM, we used keras-rl \cite{plappert2016kerasrl}, TensorFlow 1.7.0 and Keras 2.1.5 and these Python interfaces. We also used NumPy package (version 1.14.2) for RWR computation with 28 hardware threads. Figure \ref{fig:stack} represents the software stack of IGPM and PEM.

\subsection{Computing Environment}
We conducted all performance evaluations on one of parallel and distributed compute nodes in TSUBAME 3.0 supercomputer in Tokyo Institute of Technology. Each compute node has Intel Xeon E5-2680 V4 Processor CPUs (Broadwell-EP, 2.4GHz, 14 cores, 2 sockets), 4 Tesla P100 for NVLink-Optimized Server GPUs and 256GB (DDR4-2400 32GBx8) RAM.

\subsection{Datasets and Scenario}
We used temporal graph data from Harvard Dataverse \cite{V6AJRV_2017} and SNAP Network datasets \cite{snapnets}. Table \ref{tb:dataset} is the list of temporal graph data we used in the experiments. The column ``Steps'' means the number of graph structure updates when edges are added. In the graph data ``friends2008'', we extracted all friendship edges created in 2008 from friends network \cite{V6AJRV_2017}, and set the step value from the hour of each timestamp. In other graph data, we extracted all edges with timestamps and set the step value from the day.

\begin{table}[htbp]
\centering
\begin{tabular}{|l|r|r|r|r|}
 \hline
 Name & Category & Vertices & Edges & Steps \\ \hline
 friends2008 \cite{V6AJRV_2017} & Social & 224,879 & 3,871,909 & 6,893 \\ \hline
 transactions \cite{V6AJRV_2017} & Social & 112,130 & 538,597 & 1,779 \\ \hline
 sx-askubuntu \cite{snapnets} & Webpage & 159,316 & 964,437 & 2,060 \\ \hline 
 sx-mathoverflow \cite{snapnets} & Webpage & 24,818 & 506,550 & 2,350 \\ \hline
\end{tabular}
\caption{Temporal Graph Data Sets Used in Our Experiments}
\label{tb:dataset}
\end{table}

As query pattern graphs, we used the following type of small graphs: triangle, square, star and complete graph.
\begin{enumerate}
    \item Triangle (loop graphs with 3 vertices)
    \item Square (loop graphs with 4 vertices)
    \item Star query graph with 5 vertices (4 leaf vertices)
    \item Complete graph with 4 vertices and 6 edges
\end{enumerate}

In the reinforcement learning component of PEM, we used \textit{Epsilon-greedy} policy as the decision making policy from unit values of the output layer of the DQN. This policy selects an action corresponding to the unit with the highest state-action value(greedy action) with probability $1-\epsilon$, and it chooses an action uniformly at random with probability $\epsilon$. In the following experiments using reinforcement learning, we set the value of $\epsilon$ to 0.5.

We measured elapsed time and the number of re-computed vertices for 10 steps, from 101 steps to 110 steps. We started the measurements of them after edge additions for 100 steps because the initial state of input data graphs do not have enough edges and too sparse to extract many patterns.

In the following experiments, we compared the three graph pattern matching algorithms: batch (re-compute graph pattern matching process from scratch for each step), naive incremental (incremental pattern matching using results of the previous step, with the fixed community size) and adaptive incremental (incremental pattern matching with adaptive community size).

\subsection{Performance of IGPM}

\begin{figure}[htb]
    \centering
    \includegraphics[width=0.75\columnwidth]{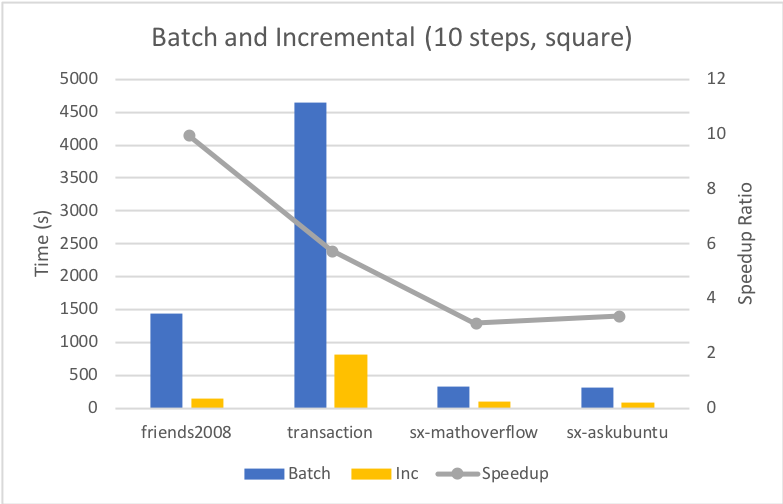}
    \caption{Elapsed time and speedup ratio of the batch and the naive incremental graph pattern matching with a square query}
    \label{fig:batch_sq}
\end{figure}

Figure \ref{fig:batch_sq} shows the elapsed time of the batch (Batch) and naive incremental (Inc) graph pattern matching with the square query pattern graph. The highest speedup ratio against the batch processing is 9.98 in friends2008 dataset. On the other hand, the ratio in sx-mathoverflow and sx-askubuntu is 3.10 and 3.37. The efficiency of our naive incremental algorithm depends on the data graphs, but at least 3 times faster than the batch version in these datasets.

\begin{figure}[htb]
    \centering
    \includegraphics[width=0.75\columnwidth]{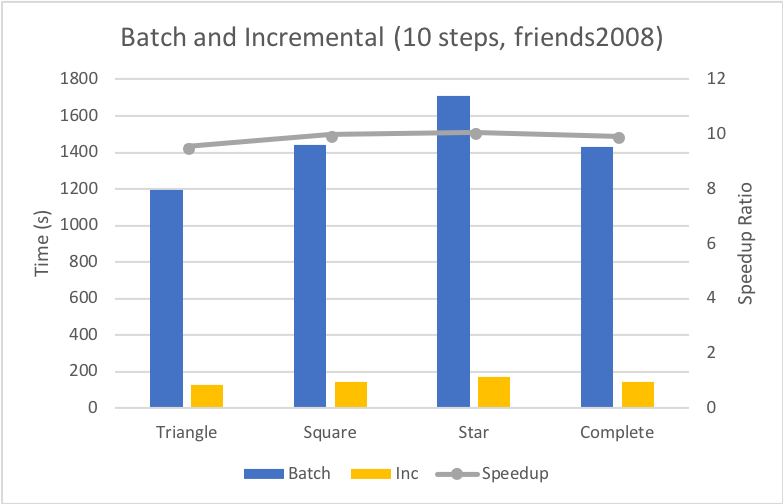}
    \caption{Elapsed time and speedup ratio of batch and naive incremental graph pattern matching with friends2008 graph}
    \label{fig:batch_fr}
\end{figure}

Figure \ref{fig:batch_fr} shows the total elapsed time and speedup ratio of the batch (Batch) and naive incremental (Inc) graph pattern matching using friends2008 dataset and four query pattern graphs. Overall, the naive incremental version is from 9.5 to 10.1 times faster than the batch version. The speedup ratio is stable in the same data graph. 

\begin{figure}[htb]
    \centering
    \includegraphics[width=0.75\columnwidth]{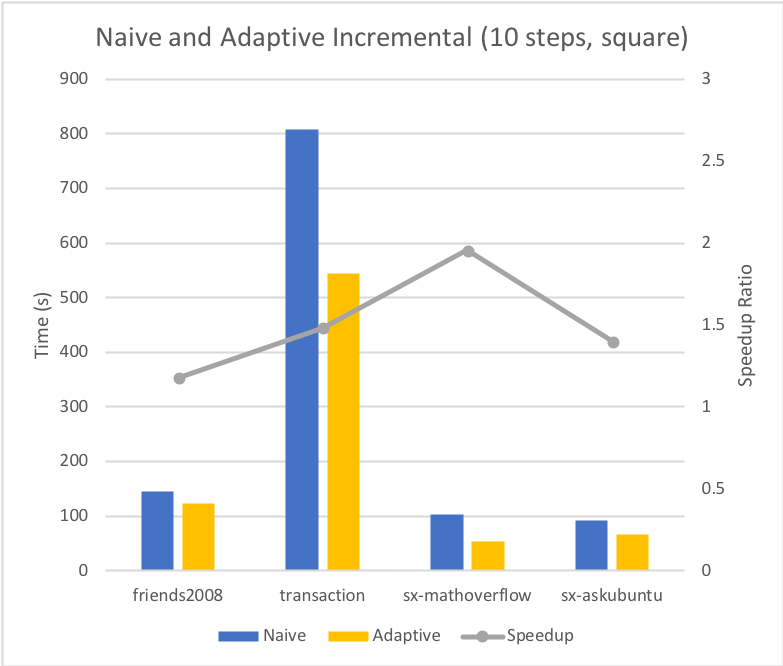}
    \caption{Elapsed time and speedup ratio of naive and adaptive incremental graph pattern matching with a square query}
    \label{fig:inc_sq}
\end{figure}

Figure \ref{fig:inc_sq} shows the total elapsed time and speedup ratio of naive and adaptive incremental graph pattern matching with the square query pattern graph. The adaptive version is from 1.17 to 1.96 times faster than the naive version.

\begin{figure}[htb]
\centering
\subfloat[friends2008]{
\includegraphics[width=0.25\textwidth]{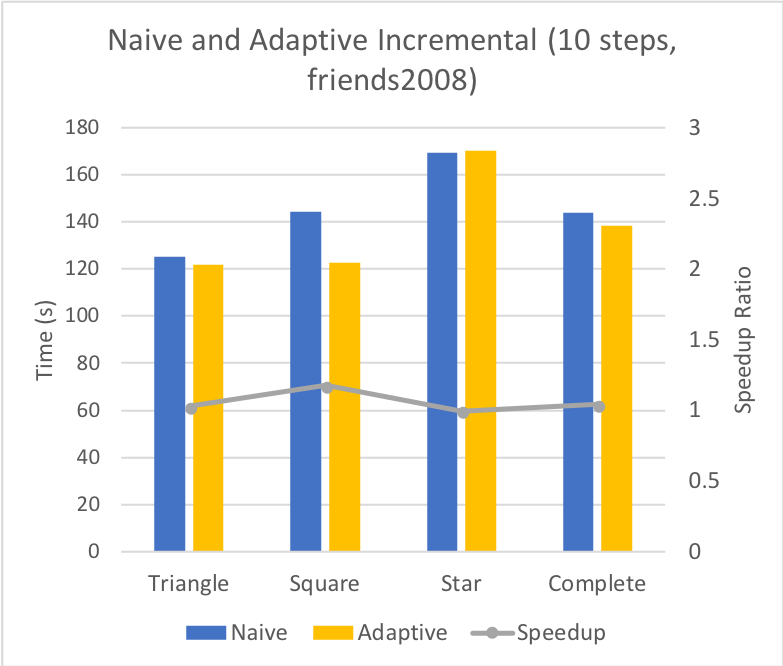}
\label{fig:inc_fr}}
\subfloat[transactions]{
\includegraphics[width=0.25\textwidth]{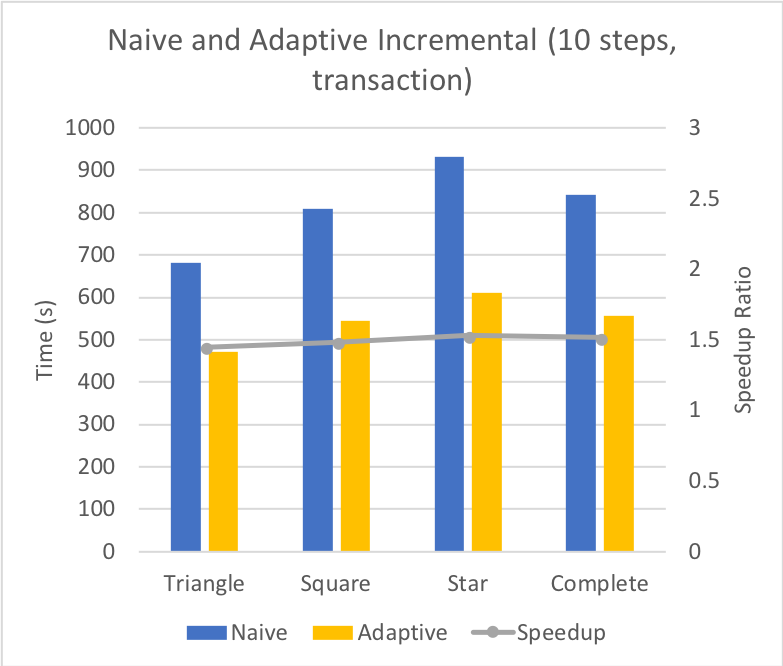}
\label{fig:inc_tx}}
\qquad
\subfloat[sx-mathoverflow]{
\includegraphics[width=0.25\textwidth]{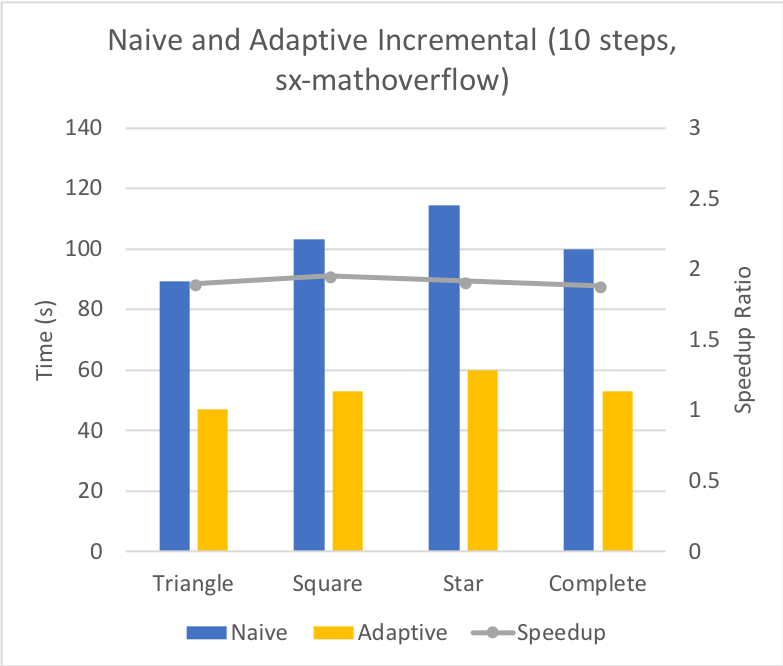}
\label{fig:inc_mo}}
\subfloat[sx-askubuntu]{
\includegraphics[width=0.25\textwidth]{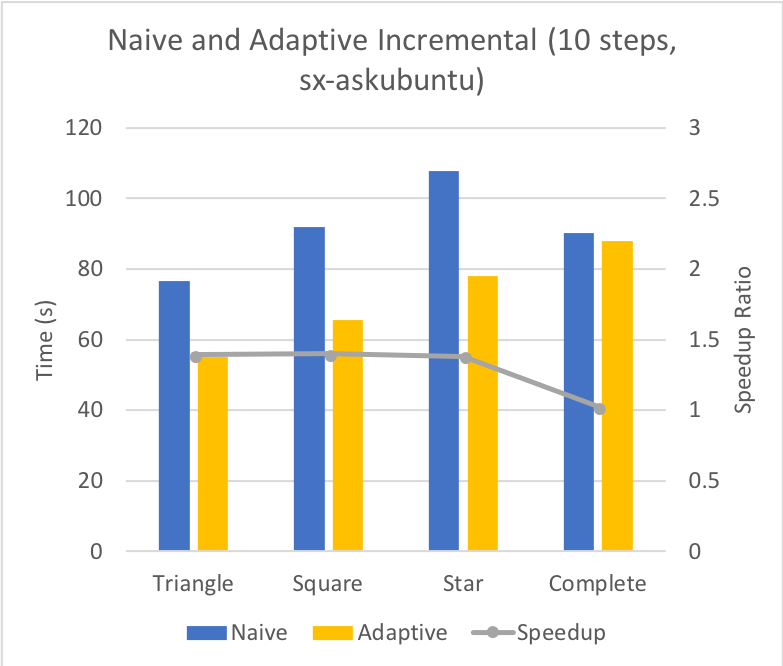}
\label{fig:inc_au}}
\caption{Elapsed time and speedup ratio of naive and adaptive incremental graph pattern matching}
\label{fig:inc_data}
\end{figure}

Figure \ref{fig:inc_fr} shows the total elapsed time and speedup ratio of naive and adaptive incremental graph pattern matching using friends2008 dataset and four query pattern graphs. The speedup ratio of adaptive version is at most 1.17.
Figure \ref{fig:inc_mo} shows the total elapsed time and speedup ratio of naive and adaptive incremental graph pattern matching using sx-mathoverflow dataset and four query pattern graphs. The speedup ratio of adaptive version is around 1.9 for all query patterns.

From the result of Figure \ref{fig:inc_data}, the efficiency of our adaptive incremental pattern matching algorithm also depends on the type of data graph rather than query patterns.


In friends2008 dataset, the total number of re-computed vertices is 37,018 in the batch version, which is 14.8 and 13.0 times as large as naive and adaptive incremental version. Because the computation time of our graph pattern matching algorithm depends on the number of vertices for re-computation, the ideal speedup ratio should be 14.8 in the incremental processing. In transactions dataset, on the other hand, the total number of re-computed vertices in the batch version is 6.97 and 9.00 times as large as that of incremental version. The speedup ratios of naive and adaptive are 5.80 and 8.08 respectively, which is better than that of friends2008 dataset.
There are several reasons the actual speedup ratio is smaller than the ideal value in the dataset. First, the RWR computation cost is still significant for dense graphs. When some vertices in large dense subgraphs are re-computed, it needs to update RWR results of all vertices in these subgraphs, which reduces the benefits of incremental computations.

\subsection{Precision Evaluation}

\begin{figure}[htb]
    \centering
    \includegraphics[width=0.75\columnwidth]{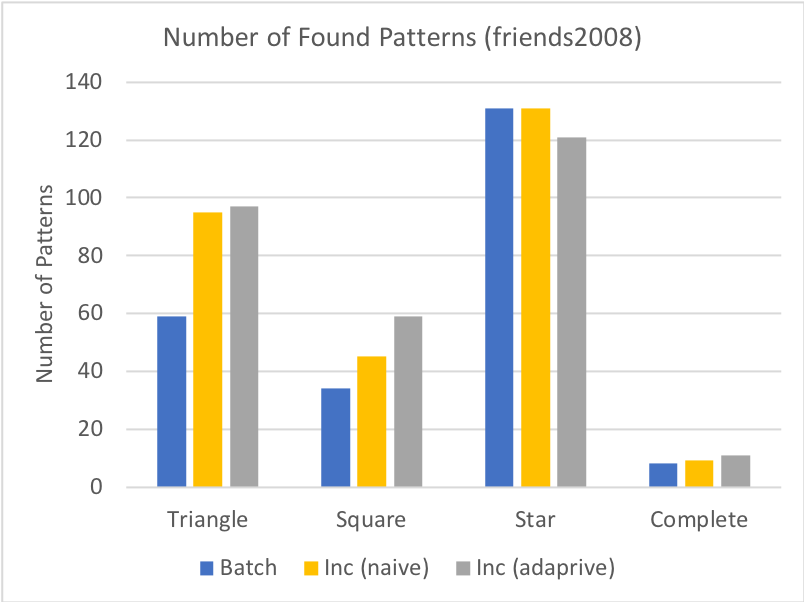}
    \caption{Number of found patterns from friends2008 graph}
    \label{fig:patterns_fr}
\end{figure}

\begin{figure}[htb]
    \centering
    \includegraphics[width=0.75\columnwidth]{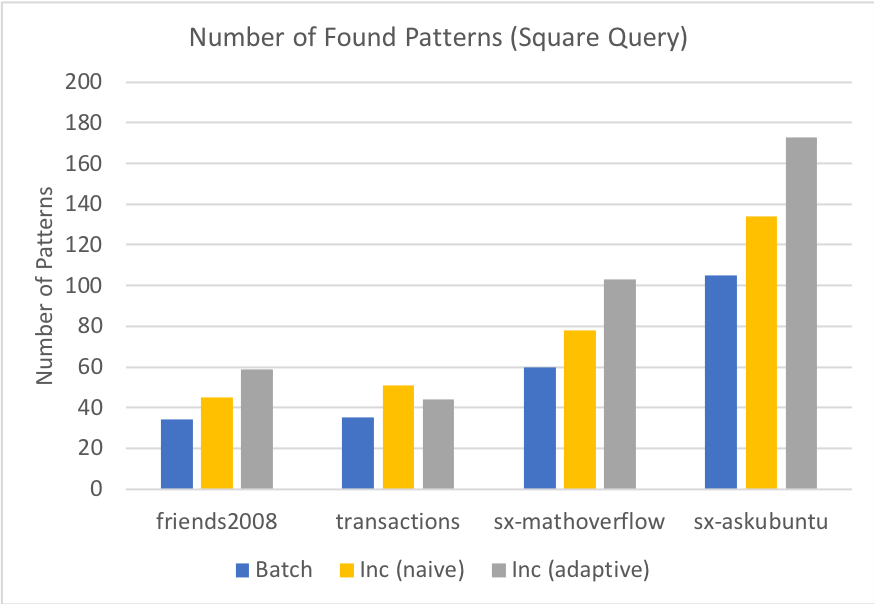}
    \caption{Number of found patterns with a square query}
    \label{fig:patterns_sq}
\end{figure}

Figure \ref{fig:patterns_fr} shows the total number of patterns correctly extracted from the friends2008 graph. Except for the star query graph, the naive and adaptive incremental version extracts from 37\% to 73\%additional patterns.
Figure \ref{fig:patterns_sq} shows the total number of patterns correctly extracted with a square query graph. The adaptive incremental version extracts 25\% to 73\% additional patterns. The reason more patterns are extracted is that our incremental algorithms re-compute all updated vertices as ``seed'' vertices and repeat the G-Ray algorithm respectively, and therefore these algorithms extracted more patterns the batch version could not find.

\section{Conclusion and Future Work}

We proposed an adaptive IGPM system with an approximate IGPM algorithm and PEM for adaptive assignment of vertices for re-computations. We also discussed the qualitative effect of our system with various input data graph type, query pattern type, and graph clustering algorithms. Then, we designed and implemented the whole system and components. Finally evaluated the efficiency of our IGPM algorithm and PEM component, and then showed up to 10.1 times speed-up in IGPM against the original G-Ray algorithm, and 10.8 times faster in our adaptive and incremental system with PEM.

We still have room for improvements of the IGPM and PEM. First, we discussed the compatibility of input data graph type and query pattern type, but we did not conduct all experiments for all combinations of these graph types. We excluded line query graphs due to the poor assumption of qualitative discussion in section \ref{sec:query}, but we will run experiments for such inefficient patterns in the future.
We could improve the reinforcement learning component of the partial execution manager. We used only the number of vertices and edges as input parameters from an input data graph, but it will the number of patterns to be extracted if we use detailed graph structure or other features of graphs. That might increase the computation cost for reinforcement learning, but it can improve the efficiency of the IGPM.

In this work, we implemented our proposed system in Python. However, it causes performance overheads and cannot handle large datasets and number of update steps than we used. As future work, we plan to implement it in C++ to make maximum use of more efficient parallel and distributed libraries such as OpenMP and MPI and computing environments such as multicore CPUs and GPUs to handle larger graph data in parallel.

\section*{Acknowledgment}
This work was partly supported by JST CREST Grant Number JPMJCR1303, Japan. This work is partially supported by the Joint Study Agreement no. W156463 under the IBM/BSC Deep Learning Center agreement.
I would like to thank the members of IBM T.J. Watson Research Center and Satoshi Matsuoka laboratory.



\bibliographystyle{IEEEtran}
\bibliography{IEEEabrv,references}
%

\end{document}